\documentclass[12pt,preprint,epsfig]{aastex}

\begin{document}
\title{Prospects for true calorimetry on Kerr black holes in core-collapse supernovae and mergers}
\author{Maurice H.P.M. van Putten$^1$, Nobuyuki Kanda$^2$, Hideyuki Tagoshi$^3$, 
        Daisuke Tatsumi$^4$, Fujimoto Masa-Katsu$^4$ and
        Massimo Della Valle$^{5,6}$}
\affil{$^1$Department of Physics, Korea Institute for Advanced Study, 87 Hoegiru, Dongdaemun-Gu, Seoul 130-722, Korea \\ 
$^2$Department of Physics, Graduate School of Science, Osaka City University, Osaka 558-8585, Japan \\ 
$^3$Department of Earth and Space Science, Graduate School of Science, Osaka City University, Osaka 558-8585, Japan \\ 
$^4$ National Astronomical Observatory of Japan, 2-21-1 Osawa, Mitaka, Tokyo 181-8588, Japan\\
$^5$Instituto Nazionale di Astrophysica - Osservatorio Astronomico di Capodimonte, Salita Moiariello 16, I-80131 Napoli, Italy\\
$^6$International Center for Relativistic Astrophysics, Piazzale della Repubblica 2, I-65122, Pescara, Italy}

\begin{abstract}
Observational evidence for black hole spin down has been found in the normalized light curves of long GRBs in the BATSE catalogue.
Over the duration $T_{90}$ of the burst, matter swept up by the central black hole is susceptible to non-axisymmetries producing 
gravitational radiation with a negative chirp. A time sliced matched filtering method is introduced to capture phase-coherence on 
intermediate timescales, $\tau$, here tested by injection of templates into experimental strain noise, $h_n(t)$. For TAMA 300, $h_n(f)\simeq 10^{-21}$ Hz$^{-\frac{1}{2}}$ at $f=1$ kHz gives a sensitivity distance for a reasonably accurate extraction of the trajectory in the time frequency domain of about $D\simeq 0.07-0.10$ Mpc for spin fown of black holes of mass $M=10-12M_\odot$ with $\tau=1$ s. Extrapolation to advanced detectors implies $D\simeq 35-50$ Mpc for $h_n(f)\simeq 2\times 10^{-24}$ Hz$^{-\frac{1}{2}}$ around 1 kHz, which will open a new window to rigorous calorimetry on Kerr black holes.
\end{abstract}

\section{Introduction}

Calorimetry and spectroscopy on all radiation channels will be key to identifying Kerr black holes in the Universe 
\citep{van02}. They may be central to some of the core-collapse supernovae (CC-SNe; \cite{woo93}) 
and mergers of neutron stars with another neutron star (NS-NS) or with a companion black hole hole (BH-NS; \cite{pac91})--the 
currently favored astronomical progenitors of cosmological gamma-ray bursts (GRBs). While all short GRB are probably produced 
by mergers, the converse need not hold in general. Black holes with a neutron star companion should have a diversity in spin 
\citep{van99}. These mixed binaries can be short and long lived, depending on the angular velocity of the black hole. 
It predicts that some of the short GRBs, produced by slowly spinning black holes, also feature X-ray afterglows \citep{van01a} 
as in GRB050509B \citep{geh05} and GRB050709 \citep{vil05,fox05,hjo05}. Long GRBs, produced by rapidly rotating black holes,
are expected to form in CC-SNe from short, intra-day period binaries (\cite{pac98,van04}), but also from mergers with a companion neutron
star \citep{van99} or out of the merger of two neutron stars \citep{bai08,van09b}. 

A diversity in the origin of long GRBs in CC-SNe and mergers naturally accounts for events with and without supernovae, notably 
GRB060614 \citep{van08a,caito09}, with and without pronounced X-ray afterglows \citep{van09} and in wind versus constant density 
host environments, that are relevant to recent studies of extraordinary {\em Swift} and {\em Fermi}-LAT events \citep{cen10,cen10b}.

Rapidly rotating black holes can sweep up surrounding matter and induce the formation of multipole mass-moments. In this process,
spin energy is catalytically converted to a long duration gravitational wave burst (GWB; \cite{van01b,van03a}). For stellar mass 
black holes, this output may be detected by advanced gravitational wave detectors for events in the local Universe. As candidate 
inner engines to long GRBs, evidence for the associated spin down of the black hole has been found in the normalized light curve 
of 600 long GRBs in the BATSE catalogue \citep{van09}. 

The high frequency range of the planned advanced detectors LIGO-Virgo \citep{bar99,arc04}, the Large-scale Cryogenic Gravitational-wave 
Telescope (LCGT, \cite{lcgt}) and the Einstein Telescope (ET, \cite{et08}) covers the quadrupole emission spectrum of orbital motions 
around stellar mass black holes, thus establishing a window to rigorously probe the inner most workings of GRBs and some of the CC-SNe.

We anticipate an event rate for long GWBs of 0.4-2 per year within a distance of 100 Mpc from the local event rate of long GRBs 
\citep{gue07}. It compares favorably with that of mergers of binary neutron stars (e.g \cite{osh08,aba10}). Since the local event 
rate of type Ib/c supernovae is $\sim 80$ per year within a distance of 100 Mpc, the branching ratio of Type Ib/c supernovae into 
long GRBs is therefore rather small, about 0.5 \citep{van04} up to $\sim 2.5\%$ \citep{gue07}. It suggests the existence of many 
failed GRB-supernovae, notably supernovae with relativistic ejecta, supernovae with pronounced aspherical explosions and radio-
loud supernovae (e.g. \cite{del10} and references therein). Therefore, the rate of events of interest to potential bursts in gravitational waves appears 
to be 1-2 orders of magnitude larger than the event rate of successful GRB-supernovae. In addition, also type II SNe, whose event 
rate is 3-4 times larger than that of type Ib/c \citep{cap99,man05}, may explode and expand asymmetrically \citep{hoe99,ish92}, 
suggesting a significant additional potential for gravitational-waves burst production.

A blind rather than a triggered search for bursts events in the local Universe seems to be appropriate by taking advantage of the 
all-sky monitoring capability of the gravitational-wave detectors in view of a beaming factor of gamma-ray bursts of $f_b<10$ 
$(\theta>25$ deg) up to a few hundred ($\theta\sim 4$ deg, \cite{fra01,van03,gue07}). A blind search is also expected to be 
competitive with current X/optical surveys for detecting the shock break-out associated with an emerging CC-SNe, as it lasts 
only a few dozens or minutes up to a few hours, and it naturally includes the possibility of long GRBs coming from merger events 
with no supernova, which may be exemplified by the long event GRB060614 of duraton 102 s discovered by {\em Swift.} 

While the energy output in long gravitational wave bursts (GWBs) produced by rapidly rotating black holes should be large, searching 
for these bursts by matched filtering is challenging in view of anticipated phase-incoherence due to turbulent magnetohydrodynamical
motions in the inner disk or torus.

Here, we focus on the detection of a trajectory in the time frequency domain produced by long GWBs, satisfying phase-coherence on short 
up to intermediate timescales. This objective goes further than the detection of a burst signal, with the aim to extract
reasonably accurate information on the burst evolution. Inevitably, the sensitivity distance for extracting trajectories
is considerably more conservative than the sensitivity distance for a detection per se. 

We shall discuss a new matched filtering detection algorithm to detect trajectories in the time frequency domain
for long GWBs with slowly varying frequencies lasting tens of seconds with intermittent phase coherence. For a burst 
lasting 50 s, 
for example, the algorithm searches by matched filtering using segmented templates on a time scale of, e.g., 1 s. This
procedure gives
a compromise between optimal matched filtering, applicable to phase-coherence extending over the entire burst duration as 
in binary coalescence of two black holes, and second order methods by correlation of independent detector signals in the 
time-domain. For our example, the compromise results in a sensitivity distance below that of optimal matched filtering 
by a factor of about $\sqrt{50}\sim 7$, and an improvement by a factor of $\sqrt{1000}\simeq$30 over second order methods 
for signals around 1 kHz.

In Section 2, we discuss the astronomical origin of long GRBs from possible both CC-SNe and mergers. In Section 3, we 
introduce a model and template for long GWBs from rapidly rotating Kerr black holes. In Section 4, we describe the 
proposed time sliced matched filtering search algorithm and the evaluation of the sensitivity distance for a 
reasonably accurate extraction of trajectories in the time frequency domain. Our findings are summarized in Section 5.

\section{Long GRBs from CC-SNe and mergers}

\begin{table}
{\bf TABLE I.} Proposed core-collapse and merger progenitors to a {\em Swift} sample of long GRBs.\\ 
\centerline{
\begin{tabular}{lrrlll}
\hline
GRB & Redshift & Duration [s] & host & constraint$^a$ & type\\
\hline
\hline
050820A$^{1,2}$   & 1.71   & 13 $\pm 2$   & UVOT $<1$ arcsec & ISM-like$^{20}$ & merger\\ 
050904$^{3,4,5}$  & 6.29   & 225 $\pm 10$ & unseen low SFR  & dense molecular cloud$^{21}$ & CC-SN \\
050911            & 0.165  & 16             & cluster EDCC 493                & no X-ray afterglow$^{6}$ & merger$^{23,24}$\\  
060418$^{7,8,9,10}$    & 1.490  & $(52\pm1)$& ISM spectrum & $\gamma$-ray efficiency$^{20}$ & merger\\
060505$^{11}$     & 0.09   &   4       & spiral, HII & no SN$^{13}$ & merger \\
060614$^{11,12,13,14}$& 0.13 & 102     & faint  SFR  & no SN$^{12,13}$ & merger$^{25,26}$ \\
070125$^{15,16}$  & 1.55   &  $>200$   & halo & ISM-like$^{22}$ & merger\\
080319B$^{17,18,19}$& 0.937&     50    & faint dwarf galaxy      & wind$^{20}$ & CC-SN\\
\hline
\end{tabular}
\label{TABLE_1}
}
$^a$ ISM-like refers to a constant host density, wind refers to a $r^{-2}$ density profile associated 
 with a massive progenitor \citep{sar98,che00}\\
$^1$\cite{ban05},$^2$\cite{fyn05};
$^3$\cite{cus06},$^4$\cite{sak05},$^5$\cite{ber07a};
$^6$\cite{ber07b};
$^7$\cite{fal06},$^8$\cite{pro07},$^9$\cite{cum06},$^{10}$\cite{cov08};
$^{11}$\cite{jak07},$^{12}$observed with a 8.2 m telescope, \cite{del06},$^{13}$observed with a 1.5 m telescope, \cite{fyn06}; 
$^{14}$\cite{geh06},$^{15}$\cite{gal06};
$^{15}$\cite{cen08},$^{16}$\cite{cha08};
$^{17}$\cite{rac08},$^{18}$\cite{tan08},$^{19}$\cite{blo09};
$^{20}$\cite{cen10},$^{21}$\cite{fra06},$^{22}$\cite{cha08};
$^{23}$\cite{pag06},$^{24}$\cite{van09};
$^{25}$\cite{van08a},$^{26}$\cite{caito09}.
\mbox{}\hskip0.01in
\end{table}

As a universal inner engine, rapidly rotating Kerr black holes can explain long GRBs from both CC-SNe and some of the mergers. They 
enable long GRBs with supernovae exclusively in star forming regions with stellar wind host environments and without supernovae such 
as GRB060614, both in and away from star forming regions including the halo such as GRB 070125. 

Some recent {\em Swift} and {\em Fermi-}LAT detections of long GRBs show events with and without X-ray afterglows such as GRB050911, 
ISM-like constant or wind-like $r^{-2}$ density profiles in the local host environments, and a diversity in distances from a local 
host galaxy, as summarized in Table I. This phenomenology is not accounted for by CC-SNe alone, and suggests that some of the long 
GRBs are associated with mergers, otherwise sharing a common long-lived inner engine.

Radiative processes around Kerr black holes are driven by frame-dragging, expressed in terms of a non-zero angular velocity $\omega$ 
of zero-angular momentum observers. It gives rise to high energy emissions along the spin axis by the induced potential energy 
$E=\Omega J_p$ for particles with angular momentum $J_p$ \citep{van09} and, contemporaneously, to a spin connection to surrounding 
matter by equivalence in topology to pulsar magnetospheres \citep{van99}. Thus, observations on high-energy emissions (in gamma-rays, 
afterglow emissions, possibly ultra-high energy cosmic rays) from leptonic jets as well as on low-energy emissions (in gravitational 
waves, MeV-neutrinos and magnetic winds, powering a supernova or a radio-burst) emanating from matter is required for full calorimetry. 

According to the Kerr metric, rotating black holes are energetically very similar to spinning tops in the sense of having a ratio
\begin{eqnarray}
\frac{E_{rot}}{\Omega_H J}=\frac{1}{2}\cos^{-2}(\lambda/4)=0.5-0.5858
\label{EQN_rot}
\end{eqnarray}
close to the Newtonian value of $\frac{1}{2}$ for all spin rates $\Omega_H$ associated with the angular momentum $J$ and rapidity 
$\lambda$, where $\sin\lambda = a/M,$ $a=J/M$. In the absence of a small parameter, $E_{rot}$ reaches 29\% of the total mass-energy 
of an extreme Kerr black hole, or about $1.6-6\times 10^{54}$ erg for a black hole in the mass range $4-14M_\odot$.

Existing calorimetry on GRB-afterglows points to true energies in gamma-rays $E_\gamma$ and kinetic energies $E_{KE}$ broadly 
distributed around $1\times 10^{51}$. These estimates apply to the sub-sample of GRBs for which achromatic breaks in their light 
curves have been determined, as the basis for estimating the half-opening angle of the beamed outflow by a model fit for the interaction 
with the host environment comprising about seven parameters \citep{cen10}. It has been well-recognized that there is considerable 
uncertainty in this estimation procedure, also in assuming a uniform luminosity profile across the outflow. Quite generally, the 
outflow may comprise a baryon-poor ultra-relativistic inner jet and a possibly baryon-rich and mildly relativistic collimating wind 
(e.g. \cite{ped98,fra00,kum00,ram02,van03a,ber03,hua04,pen05}), while the inner jet may be structured in different ways, e.g., with 
the highest luminosity reached at the boundary with the collimating wind \citep{van09}. Despite these uncertainties and model 
assumptions, exceptional events \citep{cen10} point to $E_\gamma\simeq 2.25\times 10^{52}$ erg (GRB060418) and 
$E_{KE}\simeq 3.56\times 10^{52}$ erg (GRB050820A), well above their typical values around $1\times 10^{51}$ erg. {\em And yet,} 
these energies represent a {\em minor} fraction of about $1\%$ of $E_{rot}$ or less following (\ref{EQN_rot}). 

The common properties in durations, spectra and true energies in the prompt gamma-ray emissions in long GRBs point to a 
universal inner engine as the outcome of the apparent diversity in astronomical origin and host environment (Table I), whose 
lifetime is intrinsic to the physical state of the energy reservoir itself. By the above, identifying it with the rapid spin 
of the central black hole leaves about 99\% of the spin energy unaccounted for and a similar fraction ``unseen" in emissions 
in gravitational waves and MeV-neutrinos, the remainder of which may be dissipated in the event horizon.

\section{Long duration gravitational-wave bursts from Kerr black holes}

The lowest order, axisymmetric component of magnetic fields in an accretion disk can form a torus magnetosphere around a black 
hole \citep{van01b}. In its lowest energy state \citep{wal74,ruf75,dok87,van01b}, a rapidly spinning black hole can hereby 
establish a spin connection to its surrounding disk or torus for catalytic conversion of a major fraction of its spin energy 
into a variety of radiation channels.

Quite generally, the torus magnetosphere will be highly time variable. When the central black hole spins rapidly, the interaction 
is governed the variance of the net poloidal magentic flux of the magnetic field to mediate energy and angular momentum transfer.
The black hole hereby sweeps up the inner face of the torus to excite non-axisymmetric instabilities, giving rise to gravitational 
radiation, MeV-neutrinos and magnetic winds. This state of suspended accretion lasts for the lifetime of rapid spin of the black hole, 
and sets the duration of a possibly accompanying GRB. During viscous spin down, a major fraction if not most of the spin energy is 
dissipated unseen in the event horizon, creating a astronomical amounts of Bekenstein entropy.

Based on the Kerr metric, the model template produced by matter swept up by a central black hole shows negative chirp with late time 
asymptotic frequency \citep{van08a}
\begin{eqnarray}
f_{GW} = 590~\mbox{Hz} \left(\frac{10M_\odot}{M}\right),
\label{EQN_f}
\end{eqnarray}
where $M$ denotes the initial mass of an initially rapidly spinning black hole. The negative chirp represents the expansion of the inner 
disk or torus during spin down of the black hole on a viscous time scale. Thus, the merger of two neutron stars features a late-time 
frequency of 1.5-2 kHz defined by the sum of the mass of the two progenitor neutron stars ($2\times 1.5-2 M_\odot$, where the
high masses refer to PSR J0751+1807 \citep{nic05} and PSR J1614-2230 \citep{dem10}). 
Lower asymptotic frequencies down to about 500 Hz are produced by 
high mass black holes formed in CC-SNe, or mergers of neutron stars with a high mass black hole companion. Mergers and CC-SNe hereby 
represent distinct astronomical progenitors of long GWBs (and of long GWBs) with a diversity in asymptotic frequencies (\ref{EQN_f}) 
corresponding to a range in the black hole mass.

The full template follows by integration of the equations of conservation of energy and angular momentum in the process of black hole 
spin down mostly against the inner face of a surrounding torus \citep{van08a}
\begin{eqnarray}
\dot{M}=-\kappa (\Omega_H-\Omega_T)\Omega_T,~~
\dot{J}=-\kappa (\Omega_H-\Omega_T),
\label{EQN_2}
\end{eqnarray}
where $\kappa$ represents the spin connection due to the (variance) in poloidal magnetic field for a black hole in its lowest energy 
state and the angular velocity $\Omega_T$ of the torus is tightly correlated to that at the ISCO. As non-axisymmetries develop 
\citep{van08a}, $\dot{M}$ and $\dot{J}$ will be carried off mostly in gravitational waves, until the fixed point $\Omega_H=\Omega_T$ 
is reached at low $a/M$. The resulting gravitational-wave templates obtain by integration of (\ref{EQN_2}) as a system of two coupled 
ordinary differential equations in response to an inital mass and spin rate. In \cite{van08a}, we calculated the model templates for 
different initial values of $a/M$. The resulting strain amplitude and frequency as seen at Earth for a source at a fiducial distance 
$D=10$ Mpc produced by a black hole with initially maximal spin $(a=M)$ is shown in Fig. 1. Here, we show the orientation averaged 
strain amplitude, which is a factor of $1/\sqrt{5}$ smaller than the amplitude at optimal orientation of the source along the line 
of sight \citep{fla98}. 

\begin{figure}
\centerline{
\includegraphics[scale=0.47]{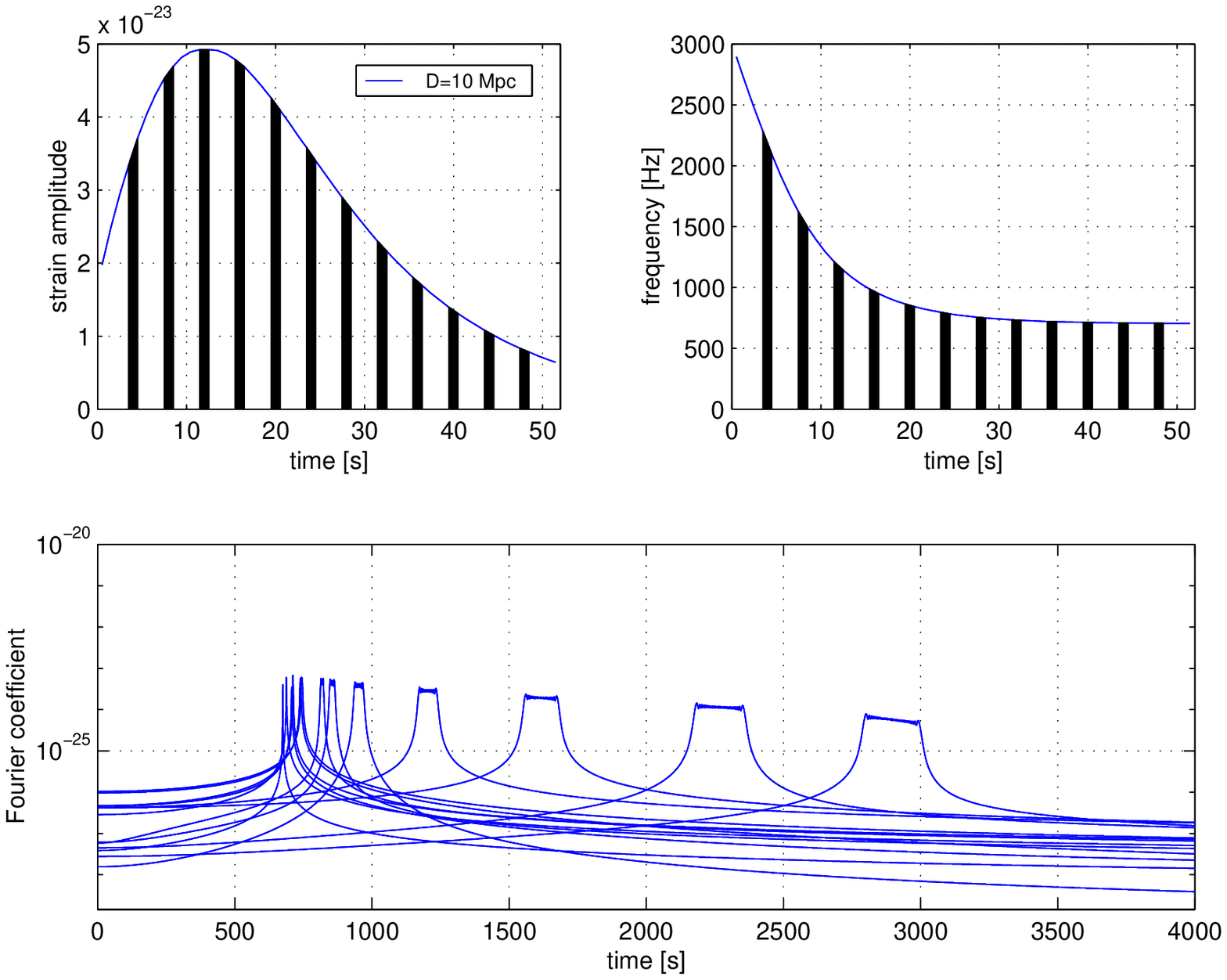}
\includegraphics[scale=0.47]{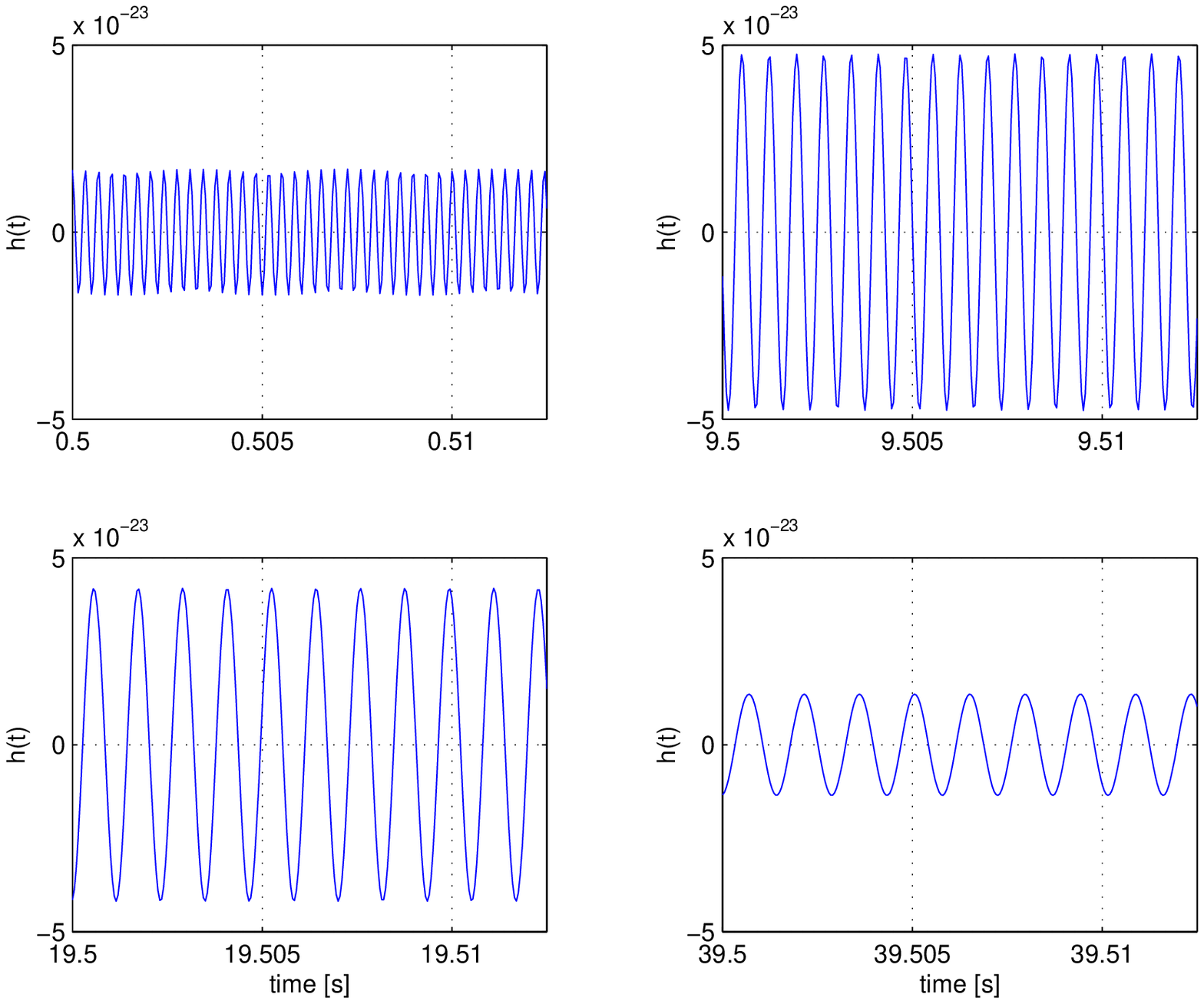}
}
\caption{($Left.$) Shown is the evolution of the orientation-averaged strain amplitude of the model template of a long GWB produced 
by a Kerr black hole of mass $10M_\odot$ at a fiducial distance of 10 Mpc ($left$ $top$) and the negative chirp ($right top$) 
with a decay in frequency ($right$~$top$) to (\ref{EQN_f}) associated expansion of the ISCO up to the fixed point $\Omega_H=\Omega_T$ 
of (\ref{EQN_2}). For time sliced matched filtering, the template is partioned into segments of intermediate duration, here sub-windows 
of 1 s duration of which a few are highlighted by the solid bars. Each segment has an effectively band-limited frequency spectrum, 
shown here for some of the 52 segments ($middle$). The maximum luminosity is reached when $a/M\simeq0.8$. For $a/M>0.8$, the luminosity 
is less than maximal as $\Omega_T\simeq \Omega_{ISCO}$ remains close to $\Omega_H$. For $a/M<0.8$, the luminosity decays, as black hole 
continues to relax to a slowly spinning, nearly Schwarzschild black hole corresponding to the fixed point $\Omega_H=\Omega_T$. 
($Right$.) Snapshots of the model templates for $h_n(t)$ in segments 1, 10, 20 and 40 out of a total of 52.}
\label{fig_1}
\end{figure}

In general, the total energy output and the frequency scale with $M$ and $M^{-1}$, respectively. The strain amplitude scales with 
$\kappa$ as a function of the mass of the torus relative to $M$ (typically about 0.1-1\%), while the total duration of the burst 
scales with $M$ and $\kappa^{-1}$.

\section{Time sliced matched filtering}

The application of matched filtering depends crucially on phase coherence in the true signal, in correlating it to a model template. 
For a magnetohydrodynamical system powered by a Kerr black hole, turbulence in the inner disk or torus inevitably creates phase 
incoherence on long time scales, inhibiting the application of matched filtering by correlating to a complete wave form template 
with the detector output. To circumvent this limitation, we first slice a model template into $N=T_{90}/\tau$ segments on intermediate 
time scales, $\tau$, for which phase-coherence may be sustained. Matched filtering is now applied using each slice, by correlating
each template slices $i$ with the detector output with arbitray offset $\delta$ in time.

\begin{figure}
\centerline{\includegraphics[scale=0.6]{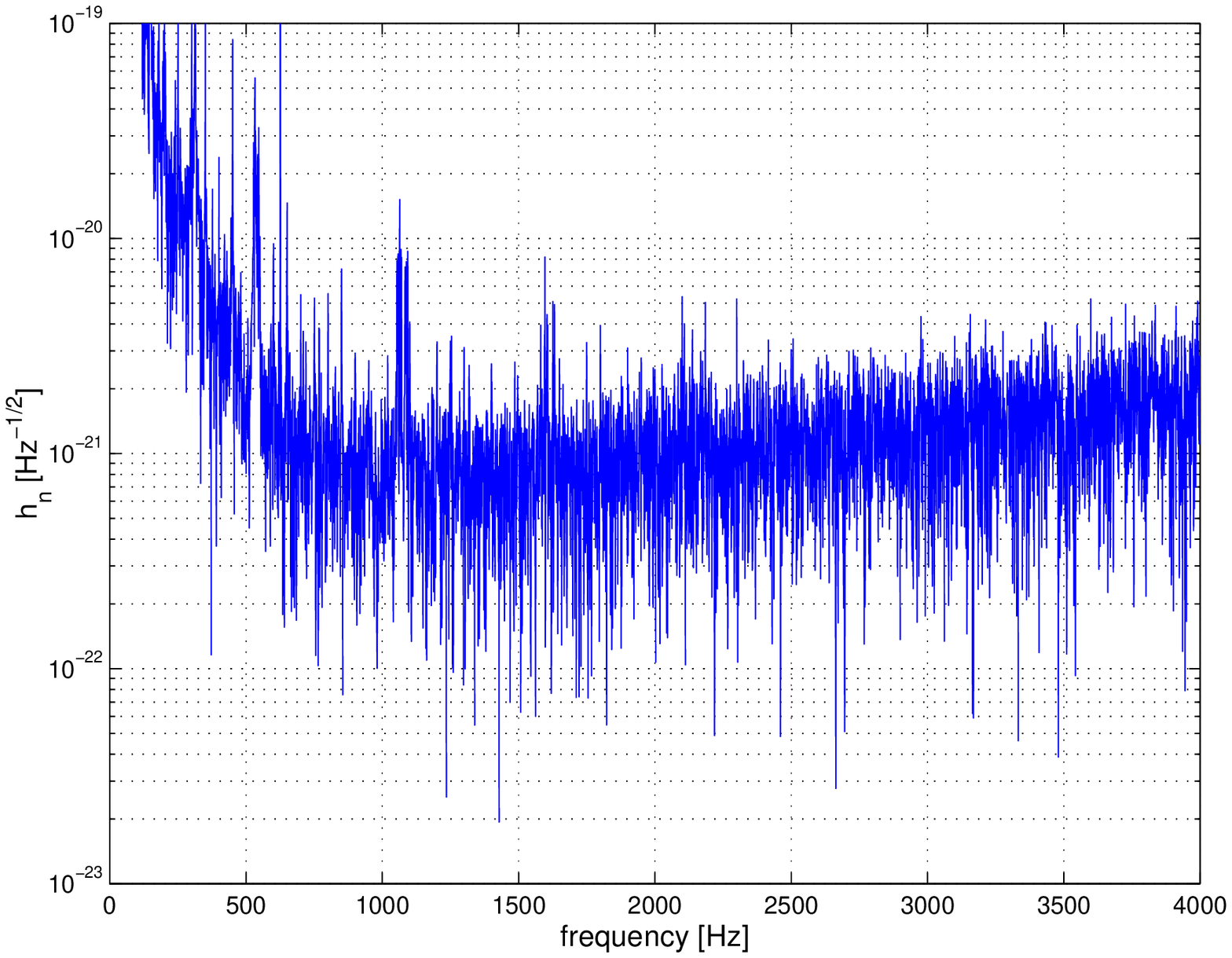}}
\centerline{\includegraphics[scale=0.6]{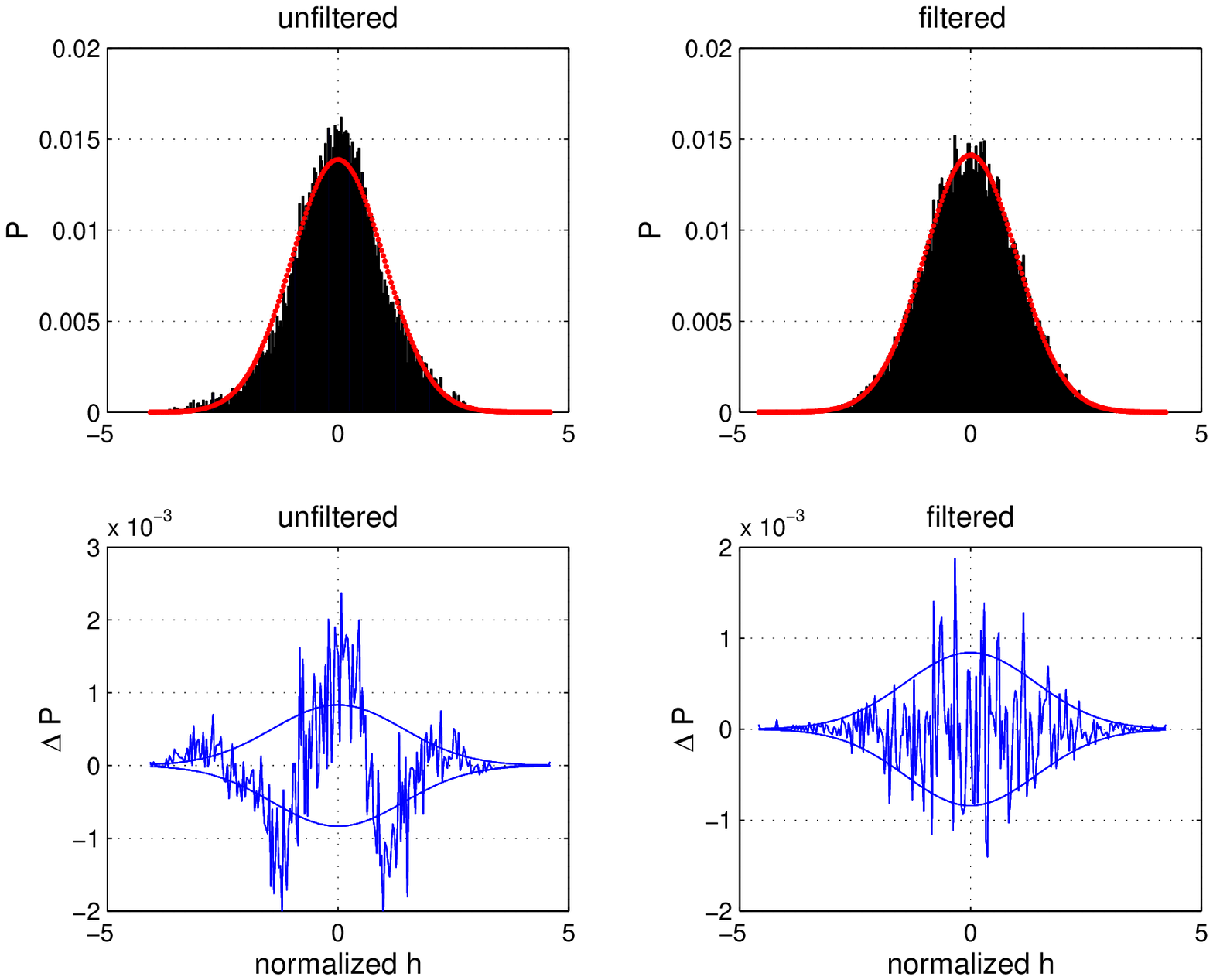}}
\caption{($Top.$) Shown is the TAMA 300 strain noise amplitude in the one-sided frequency domain in the DT9 run of 2003-4 produced by 
1 frame of 52 s duration. $(Middle.)$ The distribution is non-Gaussian largely due to low frequency noise below a few hundred Hz.
($Bottom.$) The fluctuations of the observed probability distribution of the strain noise amplitude over 1 second, 
normalized to $\sigma=1$ and plotted over 250 bins of normalized strain amplitude noise $h_n$, about to the Gaussian 
distribution with $\sigma=1$ can be compared with the expected fluctuations on the basis of the standard error in the 
mean in each bin (SEM, smooth 1-$\sigma$ curves). A band pass filter across 650 Hz - 4000 Hz recovers essentially
Gaussian noise.}
\label{fig_2}
\end{figure}

We here report on a test of our algorithm on strain noise amplitude data of the TAMA 300m detector \citep{tak04}. TAMA 300m was the 
first laser interferometric gravitational-wave detector with long duration continouus operation at a reasoble sensitivity of better 
than $10^{-20}$Hz$^{-1/2}$ in strain amplitude noise. Here, we use data from run DT8 (2/2003-4/2003) and DT9 (11/2003-1/2004) 
(partially coincident with the LIGO S3 run). The data are sampled at 20 kHz and organized in frames of $2^{20}$ samples (about 52 
seconds). 
During DT8-9, we note that some 51 CC-SNe have been observed in the local Universe with a mean redshift of 0.024 ($D= 100$ Mpc). 
Fig. 2 shows the detector strain noise amplitude for DT9. Relevant to our exploration of the detector sensitivity to the long 
GWBs is the high frequency range between a few hundred and a few thousand Hz. Given this focus and a noticeable increase in the 
detector strain noise amplitude below 550 Hz, we apply a band filter between 650 Hz and 4000 Hz to the detector strain noise 
amplitude, where the noise is Gaussian to high precision as shown in Fig. \ref{fig_2}. To excellent approximation, therefore,
the detector output $h_n(t)$ is representing by the sum of a signal $s(t)$ at the detector and white additive noise $w(t)$,
\begin{eqnarray}
h(t)=s(t)+w(t),~~s(t)=\left(\frac{10\mbox{Mpc}}{D}\right)S(t),
\label{EQN_hsw}
\end{eqnarray} 
where $D$ denotes the source distance and $S(t)$ denotes the template 
$S(t)$ $(0<t<T_{90})$ of the signal for a source at a reference distance of 10 Mpc.

For illustrative purposes, we consider a time sliced matched filtering (TSMF) with intervals $\tau=1$ s and a burst duration 
$T_{90}$ of 50 s. Slicing the template $S(t)$ $(0<t<T_{90})$ into segments of duration $\tau$ gives $N_s=T_{90}/\tau=50$ 
templates $S_i(t)$ for each slice $i$,
\begin{eqnarray}
S_i(t)=S(t_i^*+t),~~t_i^*=i\tau,~~(i=1,2,\dots,N_s,~~0<t<\tau).
\label{EQN_S}
\end{eqnarray}
Here, it is understood that the time $t$ is discrete, i.e., $t=t_j$, $t_j=(j/N_f)$ s, where $N_f=20000$ denotes the 20 kHz 
sampling rate of the TAMA 300 detector.
TSMF over segments of all data points ($N=N_s\times N_f\simeq 10^6$) in the TAMA 300 data frames starts with computing 
the Pearson coefficients
\begin{eqnarray}
\rho_i(\delta)=\frac{(S_i\cdot h)_{\delta}}{||S_i|| ~||h_\delta||}
\label{EQN_rho}
\end{eqnarray}
as a function of the unknown offset $\delta=i/20000$ ($i=0,1,\dots, N$) over a complete frame of $N$ samples obtained
at the sampling rate of 20 kHz, representing the uncertainty in continuation of phase between the time slices, where
\begin{eqnarray}
(S_i\cdot h)_{\delta} = \Sigma_{0<t_j<\tau} S_i(t_i^*+t_j)h(\delta+t_j)
\end{eqnarray}
denotes the discrete inner product of the $S_i(t_i^*+t_j)$ and the $h(\delta+t_j)$ with $L_2$ norms
\begin{eqnarray}
||S_i||      = \left[\Sigma_{0<t_j<\tau} S^2(t_i^*+t_j)\right]^{1/2},~~
||h_\delta|| = \left[\Sigma_{0<t_j<\tau} h^2(\delta+t_j)\right]^{1/2}.
\end{eqnarray}
Thus, $\rho_i(\delta)$ represents the cosine between the $S_i$ and $h$ in (\ref{EQN_rho}).

To define a signal to noise ratio $SNR_i$ for each slice $i$, we proceed with the normalized Pearson coefficients
\begin{eqnarray}
\hat{\rho}_i(\delta) = \frac{\rho_i(\delta)}{\sigma_i},
\end{eqnarray}
where $\sigma_i$ denotes the standard deviation of $\rho_i(\delta)$ as a function of $\delta$,
and define 
\begin{eqnarray}
SNR_i = \max_\delta \hat{\rho}_i(\delta).
\end{eqnarray}

In order to extract a trajectory in the time frequency domain, we 
proceed with correlations $\hat{\rho}_i(\delta)>4$, and select the
largest correlation in each time slice. This filter extracts at most 52 data points from each frame of 52
seconds when $\tau=1$. 

Fig. 3 shows the statistical results over 10 frames for spin down of a
black hole of $M=10M_\odot$ at distances of $D=0.050$ Mpc and $D=0.070$ Mpc. Here, extraction of a 
time-frequency trajectory is performed by polynomial interpolation of the aforementioned $\le 52$ filtered
points with rejection of points with large scatter. Fig. 4 indicates a sensitivity distance of about 
$D\simeq 0.07-0.1$ Mpc in the black hole mass range $M=10-12M_\odot$.

\begin{figure}
\centerline{\includegraphics[scale=0.45]{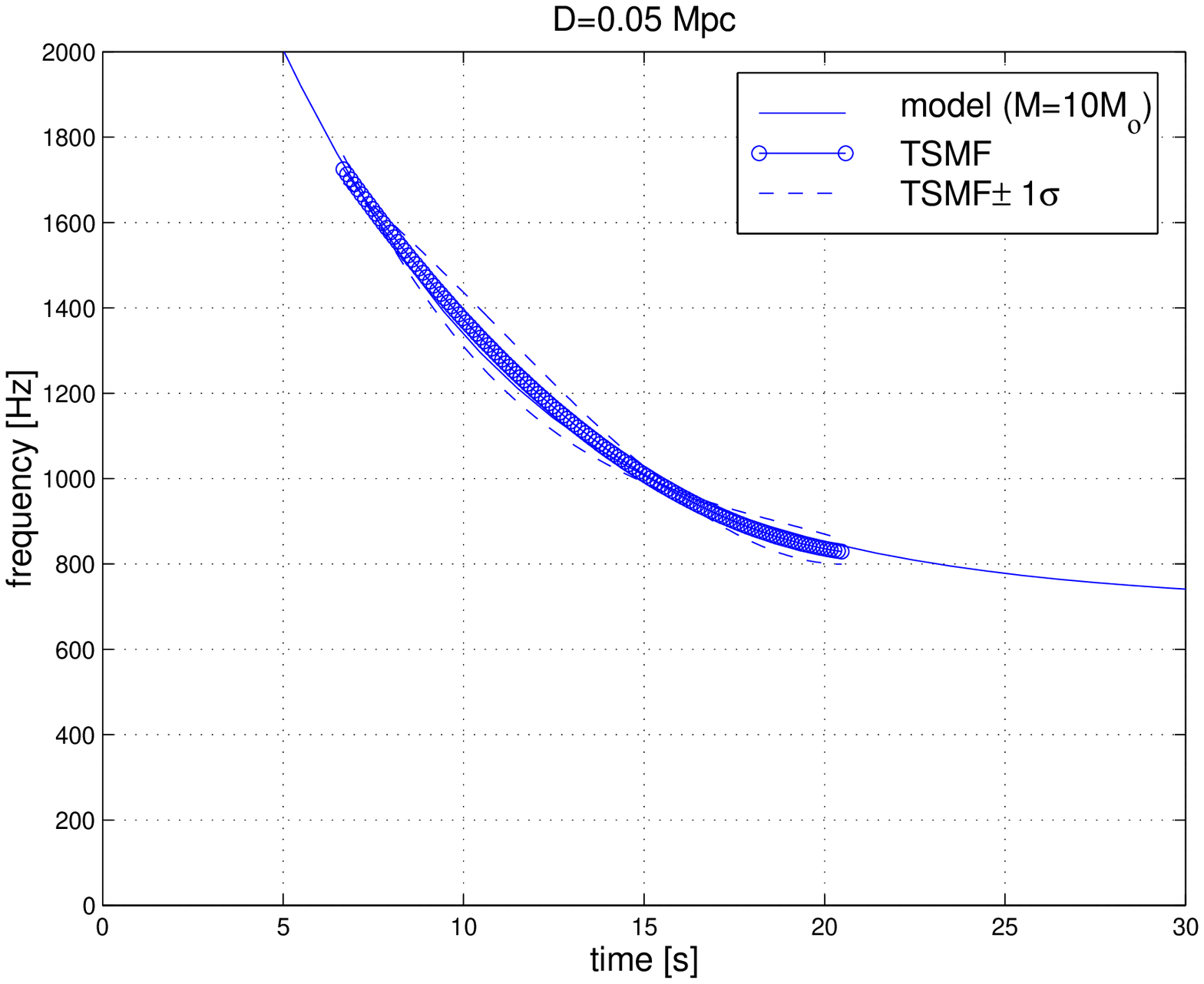}\includegraphics[scale=0.45]{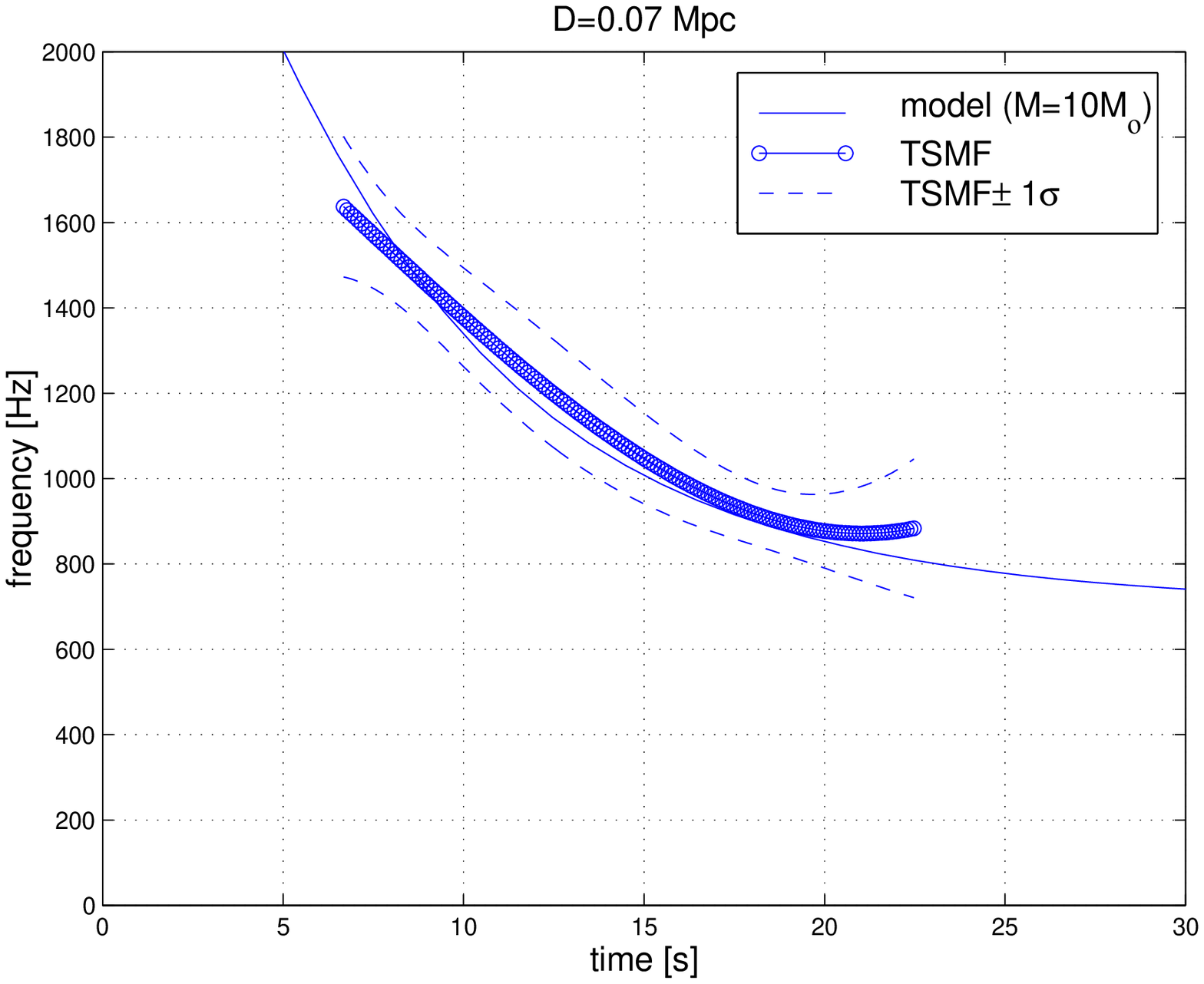}}
\centerline{\includegraphics[scale=0.45]{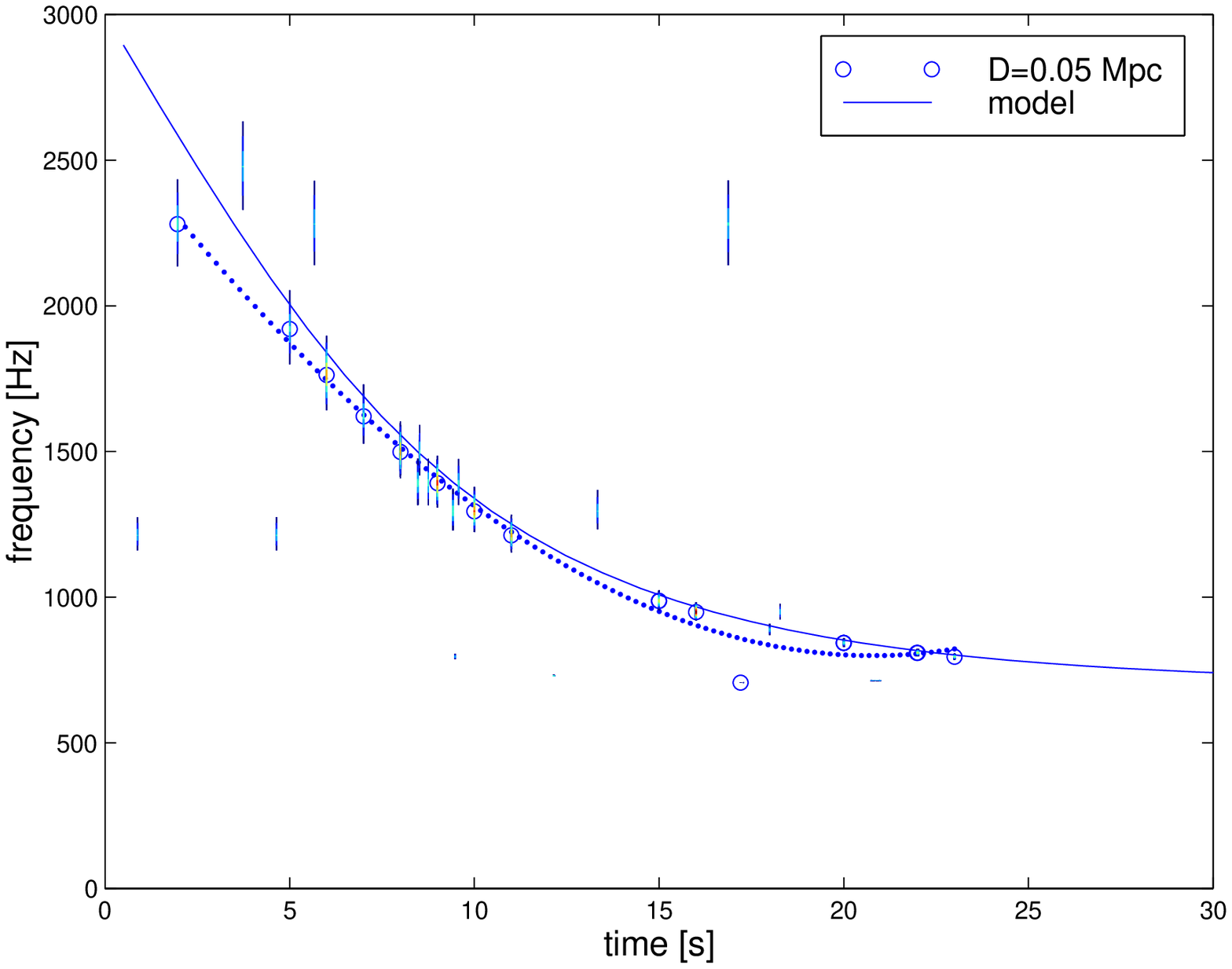}\includegraphics[scale=0.45]{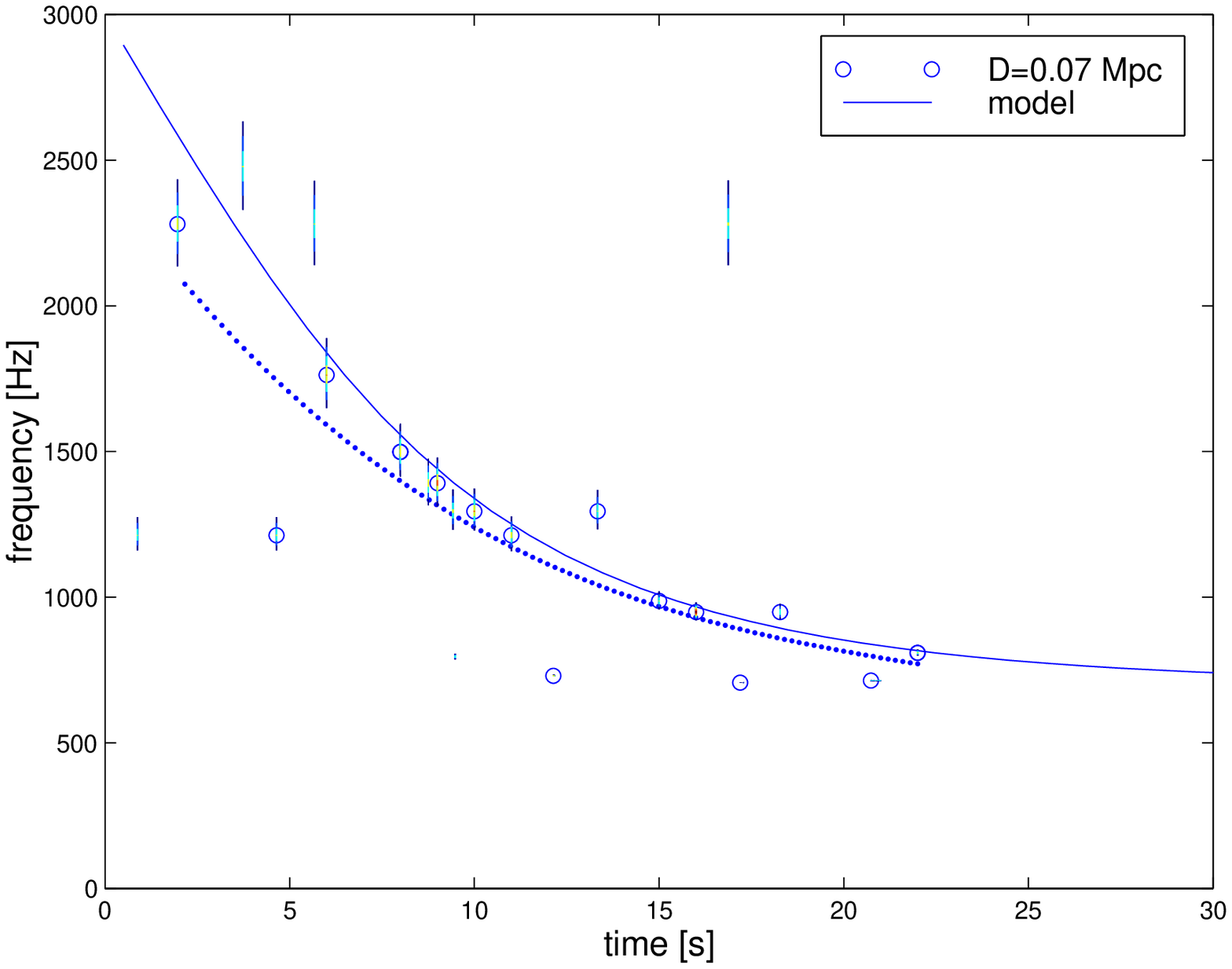}}
\centerline{\includegraphics[scale=0.45]{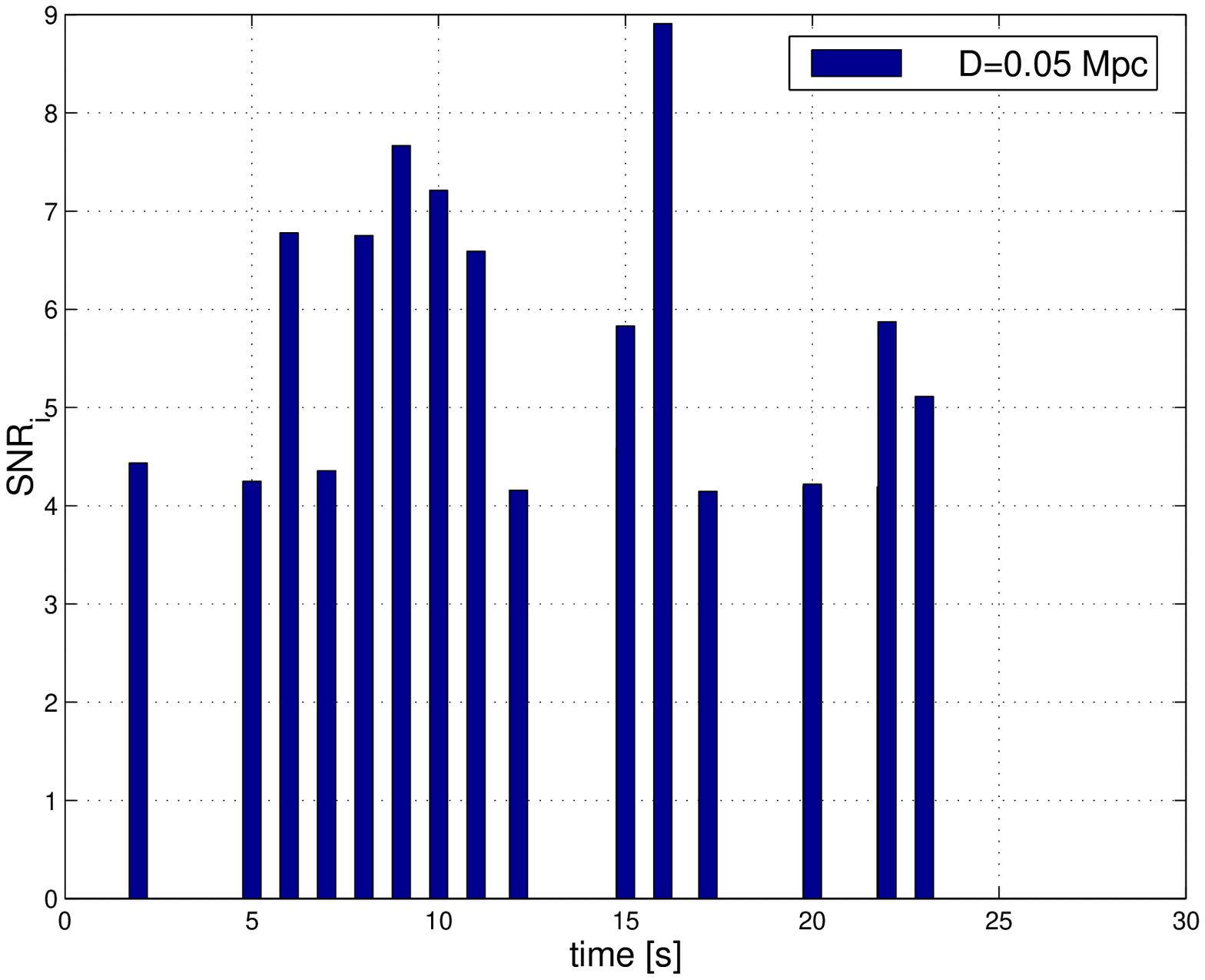}\includegraphics[scale=0.45]{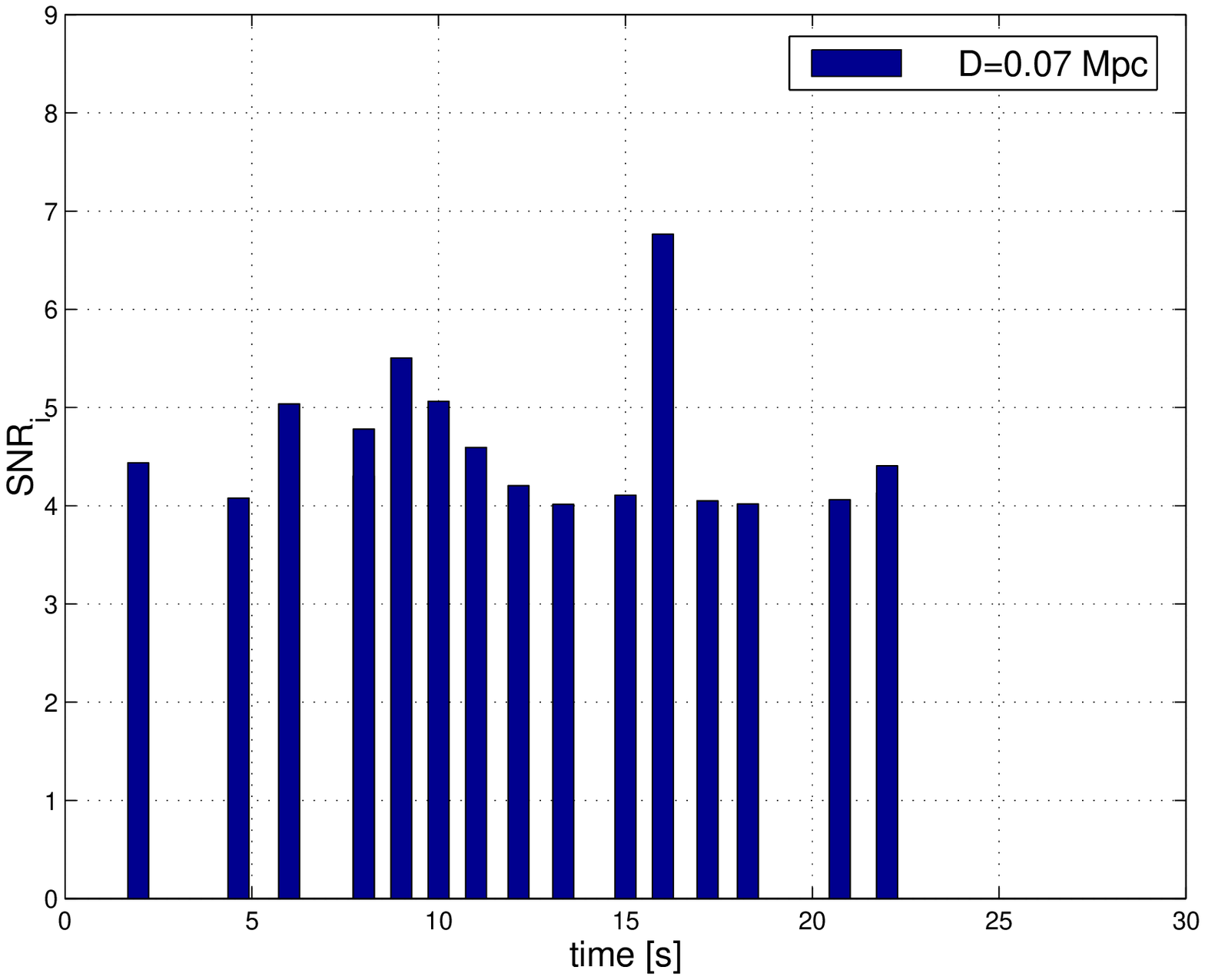}}
\caption{($Top.$) Time frequency trajectories in the time-frequency domain extracted by time sliced matched filtering (TSMF) 
for spin down of a $M=10 M_\odot$ black hole to a remnant of $8.4 M_\odot$. The results shown in the top windows represent 
the outcome of injections into 10 TAMA 300 frames of 52 second durations for two source distances. ($Middle$.) The extraction 
is based on the largest SNR$>4$ in $\rho_i(\delta)$ (on $0<\delta<23$ s, containing most of the signal) and at most one data 
point in each of the $\tau = 1$ s bins $i=1,\dots,51$ ($vertical~lines$), followed by a polynomial interpolation ($dotted$ $curve$) 
of points with reduced scatter ($circles$). ($Bottom.$) The signal to noise ratios of the data points used in the polynomial fit.}
\label{fig_4}
\end{figure}
\begin{figure}
\centerline{\includegraphics{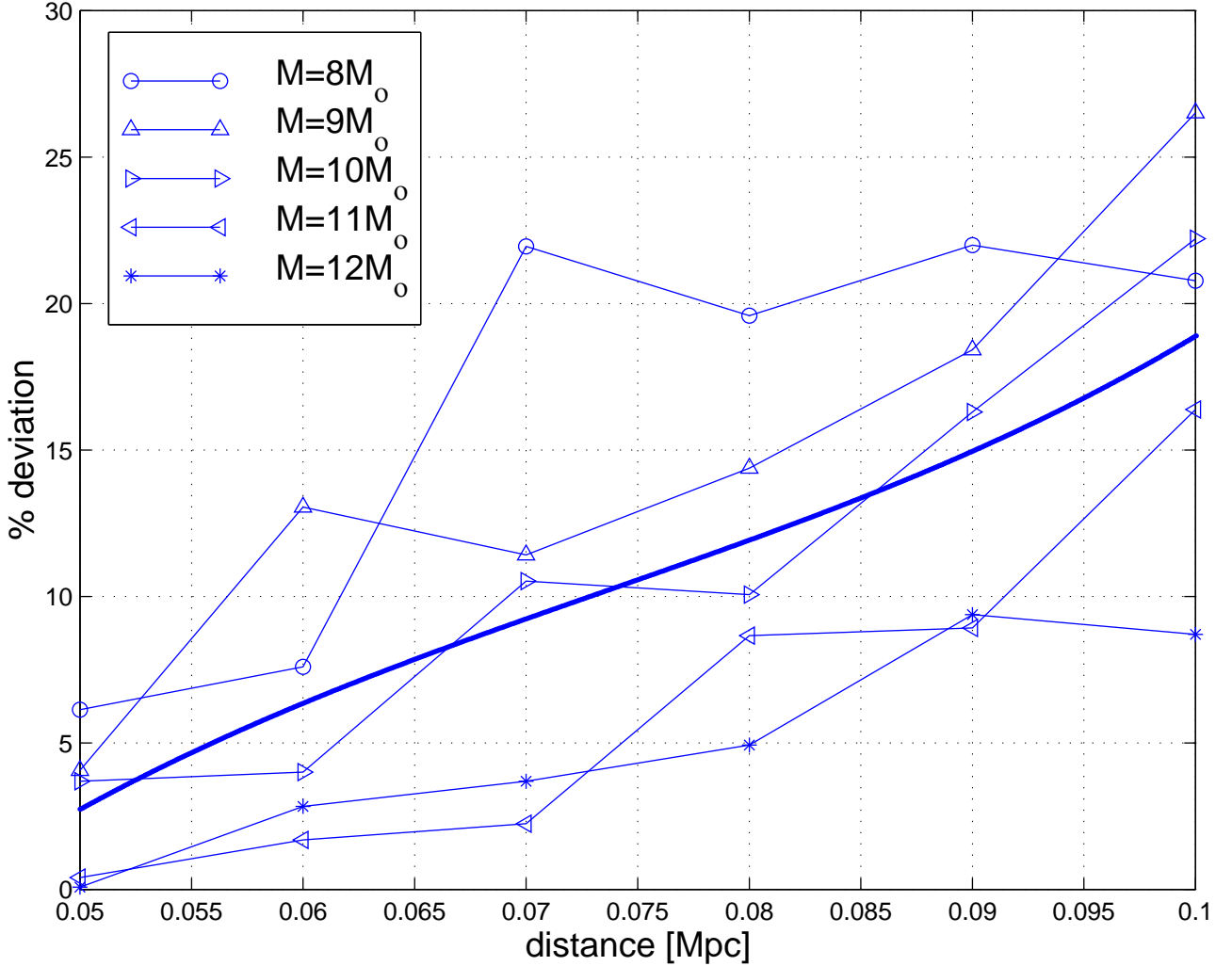}}
\caption{Shown is the resulting deviation of the extracted time-frequency trajectory from the model trajectory of the
injected signal as a function of distance $D=0.05-0.1 $Mpc and black hole masses $M=8-12 M_\odot$. The curves 
represent the standard deviation of the deviations in the extracted trajectories over a common time interval
$12-16$ s. The results show sensitivity distances from $D=0.05$ Mpc for $M=8M_\odot$ to about
$D=0.1$ Mpc for $M=12 M_\odot$ and the average trend line ($thick$ $line$).}
\label{fig_5}
\end{figure}

\begin{table}
{\bf TABLE II.} Estimated $1-\sigma$ uncertainty in the extracted time frequency trajectories by TSMF as a function of 
distance applied to long bursts in GWs produced by black hole spin down against high density matter, expected to form 
in some of the CC-SNe and mergers of neutron stars with a rapidly rotating companion black hole. We do not include 
results on neutron star-neutron star mergers, in view of their relatively high frequencies of 
1.5-2 kHz away from the region of maximal sensitivity of the existing gravitational wave detectors.
\centerline{
\begin{tabular}{ccccccc}
\hline
Mass ($M_\odot$) & $D$[TAMA]$^{a}$ (Mpc) & $D$[Adv]$^{b}$ (Mpc) & R$_D^{c}$  & $\sigma$ & max(SNR$_i$)$^{d}$ & sum(SNR$_i$)$^{e}$\\
\hline
\hline
8  & 0.05   & 25 & 0.1 & 6\%    & 6.4  & 74 \\
8  & 0.07   & 35 & 0.3 & 22\%   & 5.0  & 54 \\
8  & 0.10   & 50 & 1.2 & 27\%   & 4.5  & 46 \\  
10 & 0.05   & 25 & 0.1 & 4\%    & 8.2  & 96 \\
10 & 0.07   & 35 & 0.3 & 11\%   & 6.2  & 68 \\
10 & 0.10   & 50 & 1.2 & 22\%   & 4.9  & 43 \\
12 & 0.05   & 25 & 0.1 & $<$1\% & 10.5 & 130\\
12 & 0.07   & 35 & 0.3 & 4\%    & 7.8  & 87 \\
12 & 0.10   & 50 & 1.2 & 9\%    & 6.0  & 61 \\
\hline
\end{tabular}
\label{TABLE_2}
}
$^a$ With $h_n\simeq 10^{-21}$ Hz$^{-\frac{1}{2}}$ at 1 kHz during DT8 (2/2003-4/2003). \\
$^b$ With $h_n=2\times 10^{-24}$ Hz$^{-\frac{1}{2}}$ at 1 kHz.\\
$^c$ Estimated event rate within distance $D$[Adv] assuming 10 times more relativistic CC-SNe than successful GRB-SNe with
otherwise the similar inner engines and the observed event rate of 1 long GRB per year within $D=100$ Mpc.\\
$^d$ Based on $\rho_i(\delta)$, $0<\delta<23$ s, $\tau = 1$ s and averages over 10 frames.\\
$^e$ Based on SNR$_i>4$ and averages over 10 frames.\\
\mbox{}\hskip0.01in
\end{table}

\section{Conclusions and outlook}

The diversity in multi-wavelength phenomenology on long GRBs strongly suggests a common inner engine that is intrinsically long-lived 
representing the outcome of various astronomical scenarios. We here identify the inner engine with rapidly rotating Kerr black holes,
whose lifetime is set by the secular time scale of spin in a process of spin down against surrounding high density matter.  At present,
the most quantitative observational evidence for black hole spin down is found by matched filtering analysis of 600 light curves of 
long GRBs in the BATSE catalogue \citep{van09}.
This mechanism points to major contemporaneous emissions in gravitational waves and MeV-neutron emissions, that is likely to be dominant 
over the presently observed electromagnetic radiation in GRB-afterglow emissions and kinetic energies in GRB-SNe. True calimetry 
on the inner engine requires observations in these non-electromagnetic windows.

Most GRBs remain unobserved, due to beaming factors of ``a few'' to about 500 \citep{fra01,van03}. For CC-SNe, there is a considerable 
uncertainty of up to several days for the time of onset based on extrapolating backwards in time the supernova optical light curve. 
Not all long GRBs appear to be associated with CC-SNe, i.e., GRB 060614 \citep{del06,fyn06,geh06,gal06} appeared without a bright supernova. 
If GRB 060614 was not an anomalously faint (``failed") supernova, it may have been a merger event \citep{van08a,caito09}. Similar considerations
apply to the halo event GRB070125, which appeared with no optically identified host galaxy \citep{cen08,cha08}. A significant fraction of binaries exist in globular clusters as indicated by their population of luminous X-ray sources (\cite{ver06} and references therein). As the number of globular clusters correlates with the luminosity of elliptical galaxies \citep{har81,bur10}, the latter in particular may be preferred sites for GRB070125 type events. Therefore, blind, untriggered searches appear to be preferred in searches for local events, exploiting the all-sky survey capability of gravitational wave detectors.

To explore the sensitivity distance of advanced gravitational-wave detectors to the anticipated long duration negative chirps from 
events in the local Universe, we introduce a time sliced matched filtering algorithm and apply it to the strain noise amplitude of
the TAMA 300 detector with signal injection. A time sliced approach can circumvent the 
limitations posed by phase-incoherence on the time scale of the duration of tbe burst, which generally inhibits optimal matched 
filtering using complete wave form templates. By injecting our model template into the strain noise data of the TAMA 300 
detector during a run with $h_n(t)\simeq 1\times 10^{-21}$ Hz$^{-\frac{1}{2}}$, we compute correlations 
between the $i-$th time slice (of duration $\tau$) and one frame (about 52 s) of detector output.

For a typical black hole mass of $10M_\odot$, the results indicate the importance of the middle and late time behavior of the 
bursts (Fig. 3) below 2000 Hz, but less so the initial spin down phase associated with higher frequencies when starting from a 
maximal rotation. This reduction results from the relatively large strain amplitude noise of the detector in the shot-noise 
region for frequencies of a few kHz. Effectively, the search is focused on the output post-maximum $(a/M<0.8)$, which obviates 
the need to consider templates over a wide range of initial spin. A complete search will include a scan over two parameters: 
a range of intermediate time scales, e.g., $\tau=0.1-1$, and scaling of frequencies of the templates to account for a diversity 
in black hole masses, e.g., between 5-15 $M_\odot$. The range $\tau=0.1-1$ s represents tens to hundreds of orbital 
periods of the torus. This time scale in coherent evolution of the torus might correspond to that of sub-bursts
in the light curves of long GRBs (e.g. \cite{van99}), and quasi-steady evolution of the torus on a time scale of at least tens 
of orbital periods follows from an upper bound of about 10\% on the electromagnetic field energy it can support relative to its
kinetic energy \citep{van03a}.

Our estimated TAMA 300 sensitivity distance for extracting time frequency trajectories is summarized in Table II. The results show
$D\simeq 0.070-0.10$ Mpc for black holes in the mass range $M=10-12 M_\odot$, which compares favorably with the sensitivity distance 
for neutron star-neutron star coalescence. Extrapolation points to a sensitivity distance $D\simeq 35-50$ Mpc for a strain noise amplitude 
of $2\times 10^{-24}$ Hz$^{-\frac{1}{2}}$ in the planned advanced detectors LIGO-Virgo and the LCGT. This sensitivity distance serves 
as a conservative estimate for the sensitivity distance for a detection. In particular, the sums of the 15 SNRi's shown in Fig. 4 are 
95 and 77 for $D=0.05$ Mpc and, respectively, $D=0.07$ Mpc. These sums point to a sensitivity distance for a detection (with no 
particular information of behavior in the time frequency domain) on the order of a few tenths of Mpc, corresponding to sensitivity 
distance of a well over 100 Mpc for the advanced detectors. 

Based on the observed event rate of long GRBs of about 1 per year within a distance of 100 Mpc, the observable event rate suitable for extracting time frequency trajectories will depend on the abundance of the parent population of aspherical, relativistic and radio-loud Type Ib/c that may be
powered by irradiation of the stellar envelope from a long-lived black hole inner engine \citep{van03b}. It may reach one per few years if 
their event rate is about one order of magnitude larger than the rate of successful GRB-supernovae. The observable event rate could be larger 
if a fraction of the supernovae of Type II is similarly powered by long-lived black hole inner engines. A much larger sensitivity distance is 
anticipated for the planned 10 km ET in Europe. Conceivably, all sky radio surveys, e.g., the LOw Frequency ARray \citep{lof10}, will further 
provide us with a probe for long duration radio bursts from mergers \citep{van09b} and, combined with gravitational wave surveys, provide a 
direct measurement of the relative event rate long GRBs from mergers to long GRBs from CC-SNe. 

The diversity in the origin of long GRBs in both CC-SNe and mergers \citep{van09b} and the comparible sensitivity distances
of their potential emissions in gravitational waves and those from binary coalescence suggests the need for extended searches
over the complete frequency range of both. Apart from scaling by black hole mass and a diversity in initial spin, the proposed 
long GWBs are universal, and their progenitors are revealed only by the absence or presence of a precursor signal in gravitational
waves in case of, respectively, a CC-SNe or merger event.

{\bf Acknowledgments.} The initial work for this research was partially supported by La R\'egion Centre during a visit to Le 
STUDIUM Institute for Advanced Studies/CNRS-Orl\'eans. The authors gratefully acknowledge the TAMA collaboration for providing 
the data. MVP thanks Lars Hernquist and Ramesh Narayan for stimulating discussions.

\end{document}